\documentclass[aps,twocolumn,prl,amsmath,amssymb,showpacs,superscriptaddress]{revtex4}
\usepackage{graphicx,multirow,array,fixmath}

\begin{document}
\title{Microcanonical Approach to the Simulation of First-Order Phase
Transitions.}

\author{V. Martin-Mayor} \affiliation{Departamento de F\'{\i}sica Te\'orica I,
  Facultad de Ciencias F\'{\i}sicas, Universidad Complutense, 28040 Madrid,
  Spain.}  \affiliation{Instituto de Biocomputaci\'on y F\'{\i}sica de
  Sistemas Complejos, (BIFI, Spain).}

\date{\today}
\begin{abstract}
  A generalization of the microcanonical ensemble suggests a simple strategy
  for the simulation of first order phase transitions.  At variance with
  flat-histogram methods, there is no iterative parameters optimization, nor
  long waits for tunneling between the ordered and the disordered phases.  We
  test the method in the standard benchmark: the $Q$-states Potts model
  ($Q\!=\!10$ in 2 dimensions and $Q\!=\!4$ in 3 dimensions), where we develop a
  cluster algorithm. We obtain accurate results for systems with $10^6$ spins,
  outperforming flat-histogram methods that handle up to $10^4$ spins.
\end{abstract}
 
\maketitle

Phase transitions are ubiquitous (formation of quark-gluon plasmas,
evaporation/crystallization of ordinary liquids, Cosmic Inflation,
etc.).  Most of them are of (Ehrenfest) first order~\cite{SANMIGUEL}.
Monte Carlo simulations~\cite{MONTEBOOK} are crucial for their
investigation, but difficulties arise for large system linear size,
$L$ (or space dimension, $D$). The intrinsic problem is that, at a
first order phase transition, two (or more) phase coexist. The
simulated system tunnels between pure phases by building an interface
of size $L$. The free-energy cost of such a mixed configuration is
$\Sigma L^{D-1}$ ($\Sigma$: surface tension), the interface is built
with probability $\mathrm{exp}[-\Sigma L^{D-1}]$ and the natural time
scale for the simulation grows with $L$ as $\mathrm{exp}[\Sigma
L^{D-1}]$.  This disaster is called {\em exponential} critical slowing
down (ECSD).
 
No cure is known for ECSD in canonical simulations (cluster
methods~\cite{SWENDSEN-WANG,WOLFF} do not help), which motivated the
invention of the multicanonical ensemble~\cite{MULTICANONICAL}. The
multicanonical probability for the energy density is constant, at
least in the energy gap $e^\mathrm{o}< e < e^\mathrm{d}$
($e^\mathrm{o}$ and $e^\mathrm{d}$: energy densities of the coexisting
low-temperature ordered phase and high-temperature disordered phase),
hence the name flat-histogram
methods~\cite{MULTICANONICAL,WANGLANDAU,DEPABLO,MASFLAT}.  The
canonical probability minimum in the energy gap
($\propto\mathrm{exp}[-\Sigma L^{D-1}]$) is filled by means of an
iterative parameter optimization.

In flat-histogram methods the system performs an energy random walk in
the energy gap.  The elementary step being of order $L^{-D}$ (a single
spin-flip), one naively expects a tunneling time from $e^\mathrm{o}$
to $e^\mathrm{d}$ of order $L^{2D}$ spin-flips.  But the
(one-dimensional) energy random walk is not Markovian, and these
methods suffer ECSD~\cite{NEUSHAGER}. In fact, for the standard
benchmark (the $Q\!=\!10$ Potts model~\cite{WU} in $D\!=\!2$), the barrier of
$10^4$ spins was reached in 1992~\cite{MULTICANONICAL}, while the
largest simulated system (to our knowledge) had $4\times 10^4$
spins~\cite{WANGLANDAU}.

ECSD in flat histogram simulations is probably
understood~\cite{NEUSHAGER}: on its way from $e^\mathrm{d}$ to
$e^\mathrm{o}$, the system undergoes several (four in $D\!=\!2$)
``transitions''. First comes the condensation
transition~\cite{CONDENSATION,NEUSHAGER}, at a distance of order
$L^{-D/(D+1)}$ from $e^\mathrm{d}$, where a macroscopic droplet of the
ordered phase is nucleated. Decreasing $e$, the droplet grows to the
point that, for periodic boundary conditions, it reduces its surface
energy by becoming a strip~\cite{DROPLETSTRIP}, see Fig.~\ref{FIG2} (in $D\!=\!3$, the droplet becomes a cylinder, then
a slab~\cite{LUIS2}). At lower $e$ the strip becomes a droplet of {\em
disordered} phase.  Finally, at the condensation transition close to
$e^\mathrm{o}\,,$ we encounter the homogeneous ordered phase.

Here we present a method to simulate first order transitions without iterative
parameter optimization nor energy random walk.  We extend the configuration
space as in Hybrid Monte Carlo~\cite{HYBRID}: to our $N$ variables, $\sigma_i$
(named spins here, but they could be atomic positions) we add $N$ real
momenta, $p_i$.  The {\em microcanonical} ensemble for the $\{\sigma_i,p_i\}$
offers two advantages.  First, microcanonical simulations~\cite{LUSTIG} are
feasible at any value of $e$ within the gap.  Second, we obtain
Fluctuation-Dissipation Eqs.~(\ref{FD1}--\ref{FD3}) where the (inverse)
temperature $\hat \beta$, a function of $e$ and the spins, plays a role dual
to that of $e$ in the canonical ensemble.  The $e$ dependence of the mean
value $\langle\hat\beta\rangle_e$, interpolated from a grid as it is almost
constant over the gap, characterizes the transition.  We test the method in
the $Q$-states Potts model, for which we develop a cluster algorithm. We
handle systems with $10^6$ spins for $Q\!=\!10$ in $D\!=\!2$ and for $Q\!=\!4$ in $D\!=\!3$
(where multibondic simulations handle $N=10^4$~\cite{JANKE-PRIV}).


Let $U$ be the spin Hamiltonian.  Our total energy is
\begin{equation}
{\cal E}= \sum_{i=1}^{N} \frac{p_i^2}{2}\ +\ U\,,\quad (e\equiv {\cal
E}/N\,,\ u\equiv U/N)\,.\label{ENERGIATOTAL}
\end{equation}
In the canonical ensemble , the $\{p_i\}$ are a trivial gaussian bath
decoupled from the spins. Note that, at inverse temperature $\beta$, one has
$\langle e\rangle_\beta= \langle u\rangle_\beta + 1/(2\beta)\,$.

Microcanonically, the entropy density, $s(e,N)$, is given by
($\sum_{\{\sigma_i\}}$: summation over spin configurations)
\begin{equation}
\mathrm{exp}[N s(e,N)]=
\int_{-\infty}^\infty\prod_{i=1}^N\mathrm{d}p_i\sum_{\{\sigma_i\}}\
\delta(Ne-\cal{E})\,,\label{MICRO1}
\end{equation}
or, integrating out the $\{p_i\}$  using Dirac's delta function,
\begin{equation}
\mathrm{exp}[N s(e,N)]
= \frac{(2\pi N)^{N/2}}{N \Gamma(N/2)}\,\sum_{\{\sigma_i\}}\, (e-u)^{\frac{N}{2}-1} \theta(e-u)\,.\label{MICRO2}
\end{equation}
The Heaviside step function, $\theta(e-u)$, enforces $e>u$.
The microcanonical average at fixed $e$ of a generic function of $e$ and the
spins, $O(e,\{\sigma_i\})$, is (see Eq.~(\ref{MICRO2}) and~\cite{LUSTIG})
\begin{equation}
\langle O\rangle_e\equiv
\frac{\sum_{\{\sigma_i\}}\,O(e,\{\sigma_i\})\,(e-u)^{\frac{N}{2}-1} 
\theta(e-u)}{\sum_{\{\sigma_i\}}(e-u)^{\frac{N}{2}-1}\theta(e-u)}\,.\label{MICROPROB}
\end{equation}
The Metropolis simulation of Eq.~(\ref{MICROPROB}), is
straightforward.

Calculating $\mathrm{d}s/\mathrm{d}e$ from Eq.(\ref{MICRO2}) we learn that
~\footnote{In~\cite{DEPABLO}, Eq.~(\ref{FD1}) was approximated as
  $\mathrm{d}s/\mathrm{d}e\approx 1/\langle 1/\hat\beta\rangle_e\,.$}
\begin{eqnarray}
\frac{\mathrm{d} s(e,N)}{\mathrm{d} e}&=&\langle
\hat\beta(e;\{\sigma_i\})\rangle_e\,,\label{FD1}\\
\hat\beta(e;\{\sigma_i\})&=&\frac{N-2}{2N (e -u)}\,.\label{BETADEF}
\end{eqnarray}
Fluctuation-Dissipation follows by derivating Eq.~(\ref{MICROPROB}):
\begin{equation}
\frac{\mathrm{d}\langle O\rangle_e}{\mathrm{d}e}=\left\langle\frac{\partial O}
{\partial e}\right\rangle_e + N\left[\langle O\hat\beta\rangle_e -
\langle O\rangle_e\langle\hat\beta\rangle_e
\right]\,.\label{FD2}
\end{equation}
As in the canonical case~\cite{REWEIGHT}, an integral version of (\ref{FD2})
allows to extrapolate $\langle O \rangle_{e'}$ from simulations at $e\geq e'$:
\begin{equation}
\langle O\rangle_{e'}=\frac{\left\langle O(e';\{\sigma_i\})\,\theta(e'-u) \left[\frac{e'-u}{e-u}\right]^{\frac{N}{2}-1}\right\rangle_e } 
{\left\langle \,\theta(e'-u)\left[\frac{e'-u}{e-u}\right]^{\frac{N}{2}-1}\right\rangle_e }\,.\label{FD3}
\end{equation}
For $e<e'$, configurations with $e< u < e'$, suppressed by a factor
$(e'-u)^{N/2 -1}$, are ignored in (\ref{FD3}).  Since we are limited in
practice to $|e-e'|\leq\sqrt{\langle u^2\rangle_e-\langle
  u\rangle_e^2}/|\mathrm{d}\langle u\rangle_e/\mathrm{d} e|\sim N^{-1/2}$, 
the
restriction $e\geq e'$ can be dropped, as it is numerically negligible.

\begin{figure}
\begin{center}
\includegraphics[width=\columnwidth,angle=0]{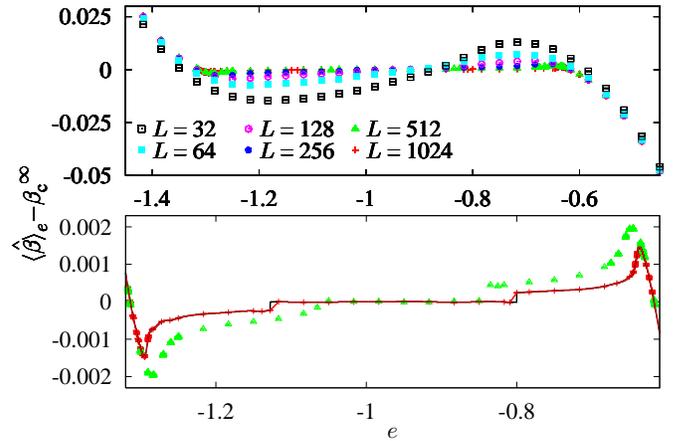}
\caption{(Color online) Excess of $\langle\hat\beta\rangle_e$ over
  $\beta_\mathrm{c}^{L=\infty}$ vs. $e$, for the $Q\!=\!10$, $D\!=\!2$
  Potts model and several system sizes.  {\bf Bottom:} magnification
  for $L\geq 512$.  The flat central region is the strip phase (the
  strip width varies at fixed surface free-energy).  Lines (shown for
  $L=1024$) are the two interpolations used for $L\geq 512$.  We
  connect 3 independent cubic splines, in the strip phase and in its
  sides, either by a linear function or by a step-like $1/100$ power.
  Differences among the two interpolations are used to estimate the
  error induced by the uncertainty in the location of the
  strip-droplet transitions.}
\label{FIG1}
\end{center}
\end{figure}

\begin{figure}
\begin{center}
\includegraphics[width=0.49\columnwidth,angle=90]{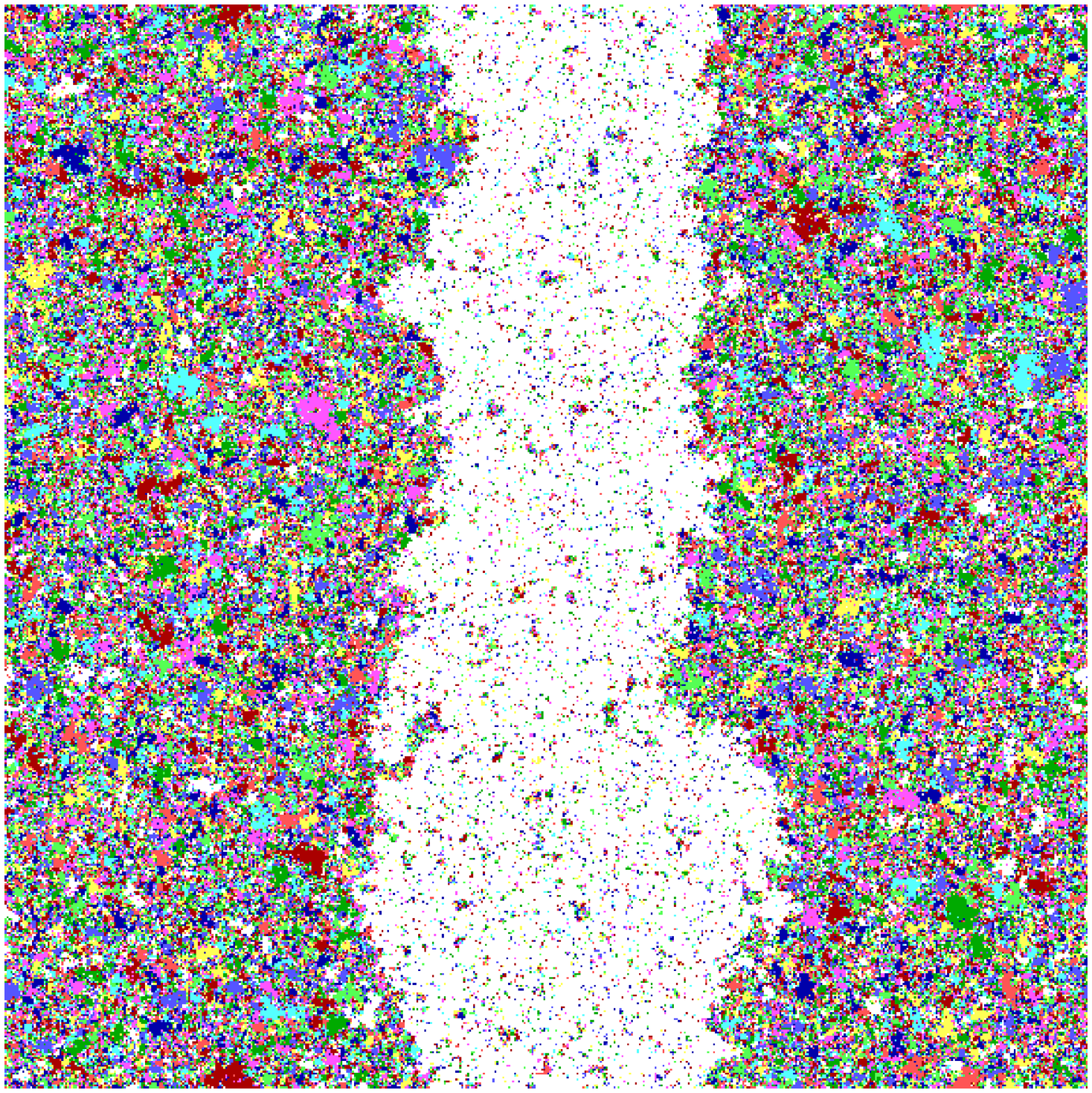}
\includegraphics[width=0.49\columnwidth,angle=90]{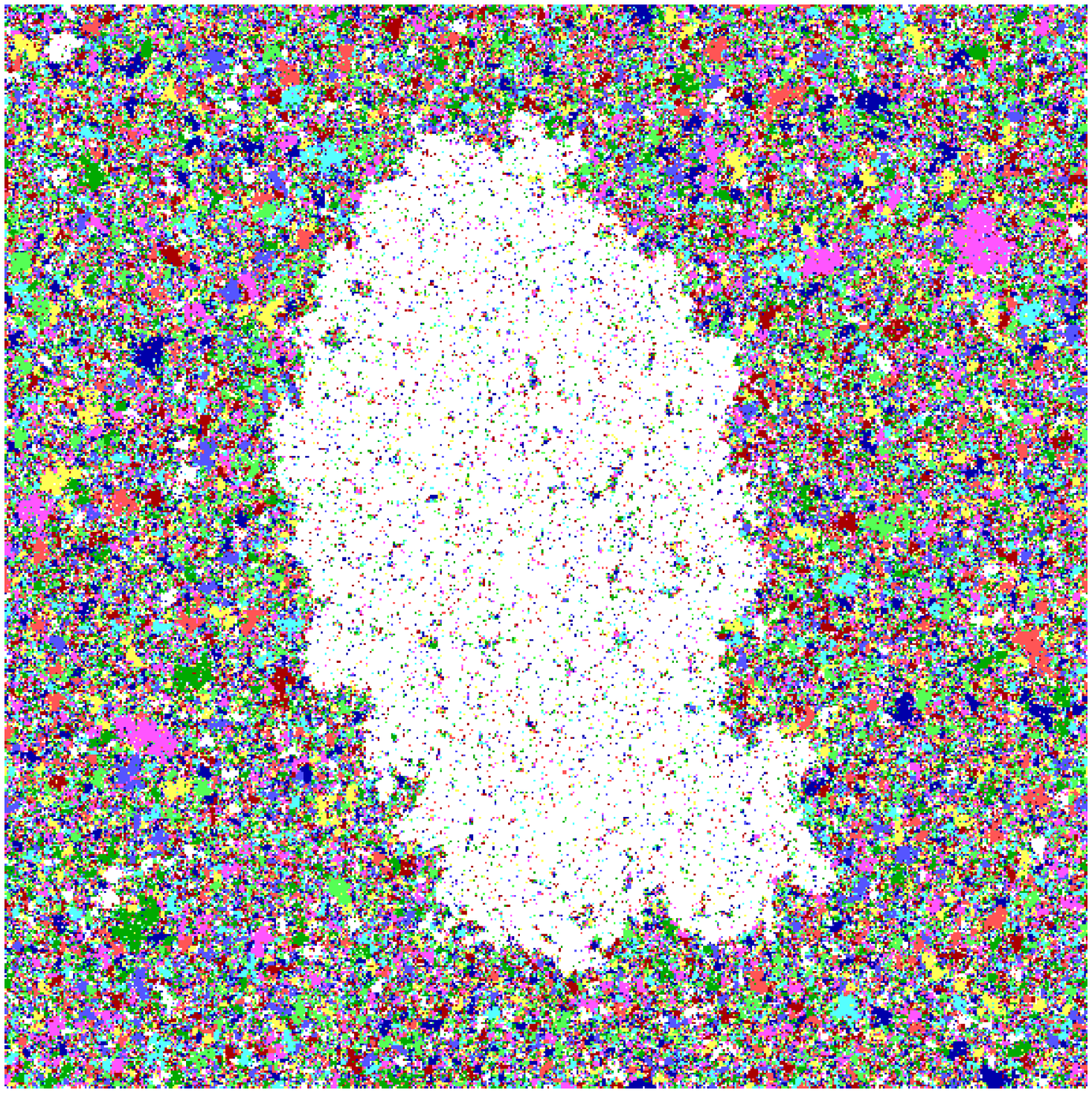}
\caption{(Color online) $L=1024$ equilibrium configurations for the
  ferromagnetic $Q\!=\!10\,,D\!=\!2$ Potts model with periodic
  boundary conditions, at the 2 sides of the droplet-strip transition,
  namely $e=-0.809$ ({\bf left}) and $e=-0.8$ ({\bf right}). }
\label{FIG2}
\end{center}
\end{figure}

The canonical probability density for $e$,
$P_{\beta}^{(L)}(e)\propto\mathrm{exp}[N(s(e,N)-\beta e)]$  follows
from $\langle\hat\beta\rangle_e$:
\begin{equation}
\log P_{\beta}^{(L)}(e_2)-\log P_{\beta}^{(L)}(e_1)=
N\int_{e_1}^{e_2}\mathrm{d}e\, \left(
\langle\hat\beta\rangle_e -\beta\right)\,.\label{LINK}
\end{equation}
In the {\em thermodynamically stable region} (i.e.
$\mathrm{d}\langle\hat\beta\rangle_e/\mathrm{d}e <0$), there is a single root
of $\langle\hat\beta\rangle_e=\beta$, at the maximum of $P_{\beta}^{(L)}$.
But, see Fig.~\ref{FIG1}, in the energy gap $\langle\hat\beta\rangle_e$ has a
maximum and a minimum ($L$-dependent spinodals~\cite{SANMIGUEL}), and there
are several roots of $\langle\hat\beta\rangle_e=\beta$. The rightmost
(leftmost) root is $e_L^\mathrm{d}(\beta)$ ($e_L^\mathrm{o}(\beta)$), a local
maximum of $P_{\beta}^{(L)}$ corresponding to the disordered (ordered) phase.
We define $e_L^*(\beta)$ as the {\em second rightmost} root of
$\langle\hat\beta\rangle_e=\beta$.

At the finite-system (inverse) critical temperature,
$\beta_\mathrm{c}^L$, one has~\cite{HEIGHT}
$P_{\beta_\mathrm{c}^L}^{(L)}(e_L^\mathrm{d}(\beta_\mathrm{c}^L))=
P_{\beta_\mathrm{c}^L}^{(L)}(e_L^\mathrm{o}(\beta_\mathrm{c}^L))$,
which is equivalent, Eq.~(\ref{LINK}) and~\cite{JANKEMIC}, to
Maxwell's construction:
\begin{equation}
0=\int_{e^\mathrm{o}_L(\beta_\mathrm{c}^L)}^{e^d_L(\beta_\mathrm{c}^L)}
\mathrm{d}e\, \left( 
\langle\hat\beta\rangle_e -\beta_\mathrm{c}^L\right)\,,\label{MAXWELL}
\end{equation}
(for large $N$, $\beta_\mathrm{c}^\infty-\beta_\mathrm{c}^L\propto
1/N\,$~\cite{BORGS-KOTECKY}). Actually, at fixed $e$ in the gap, also
$\langle\hat\beta\rangle_e$ tends to $\beta_c^\infty$ for large $N$. In the
strip phase it converges faster than $\beta_\mathrm{c}^L$, see
Table~\ref{MITABLA}.

In a cubic box the surface tension is estimated as~\footnote{In the strip
  phase (Fig.~\ref{FIG2}) {\em two} interfaces form,
  hence~\cite{BINDER}
  $P_{\beta_\mathrm{c}^L}^{(L)}(e^\mathrm{d}_L(\beta_\mathrm{c}^L))/P_{\beta_\mathrm{c}^L}^{(L)}(e^*_L(\beta_\mathrm{c}^L))
  =\mathrm{exp}[2\Sigma_L L^{D-1}]\,.$}
\begin{equation}
  \Sigma^L=\frac{N}{2L^{D-1}} \int_{e^*_L(\beta_\mathrm{c}^L)}^{e^\mathrm{d}_L(\beta_\mathrm{c}^L)} \mathrm{d}e\, \left(
    \langle\hat\beta\rangle_e -\beta_\mathrm{c}^L\right)\,.\label{SIGMAEQ}
\end{equation}
$L\to\infty$ extrapolations $\Sigma^\infty-\Sigma^L\propto 1/L$~\cite{BINDER}
are popular.

As for the specific heat, for $N\to\infty$ the inverse function of the
canonical $\langle e\rangle_\beta$ is the microcanonical
$\langle\hat\beta\rangle_e$:
\begin{equation}
\frac{\mathrm{d}\langle u\rangle_\beta}{\mathrm{d}\beta}\approx
\left[\frac{1}{2\langle\hat\beta\rangle_e^2} +
  \frac{1}{\mathrm{d}\langle\hat\beta\rangle_e/\mathrm{d} e} \right]_{e=\langle
    e\rangle_\beta}\equiv C_L(e)\,.\label{CALORESP}
\end{equation}
For large $N$, $e^\mathrm{d}_L(\beta_\mathrm{c}^L)\,$,
$e^\mathrm{o}_L(\beta_\mathrm{c}^L)\,$,
$C_L(e^\mathrm{d}_L(\beta_\mathrm{c}^L))\,$,
$C_L(e^\mathrm{d}_L(\beta_\mathrm{c}^L))\,$ tend to $e^\mathrm{d}$,
$e^\mathrm{o}$, or the specific heat of the coexisting phases (we lack
analytical hints about convergence rates).


We now specialize to the Potts model~\cite{WU}. The spins
$\sigma_i=0,1,\ldots,Q-1\,,$ live in the $N=L^D$ nodes of a (hyper)cubic
lattice of side $L$ with periodic boundary conditions, and interaction
($<ij>$: lattice nearest-neighbors)
\begin{equation}
U=-\sum_{<ij>} \delta_{\sigma_i,\sigma_j}\,.\label{UDEF}
\end{equation}
A cluster method is feasible. Let $\kappa$ be a tunable parameter and
$w(e,u,\kappa)=(e-u)^{N/2-1}\,\mathrm{exp}[\kappa Nu]\theta(e-u)$.
Our weight is $w(e,u,\kappa) \mathrm{exp}[-\kappa U]\,,$ see
(\ref{MICROPROB}), or, introducing bond occupation variables,
$n_{ij}=0,1$, and $p\equiv
1-\mathrm{exp}[\kappa]$,
\begin{equation}
  w(e,u,\kappa)\prod_{<i,j>}\left[(1-p)\delta_{n_{ij},0}\ +\ 
    p\,\delta_{n_{ij},1}\delta_{\sigma_i,\sigma_j}\right]\,,\label{MICROCLUSTER}
\end{equation}
which is the canonical statistical weight at
$\beta\!=\!\kappa$~\cite{EDWARDS-SOKAL}, but for the $\{n_{ij}\}$
independent factor $w(e,u,\kappa)$.  Hence, clusters are traced in the
standard way, but we accept a single-cluster flip~\cite{WOLFF} with
Metropolis probability $p(e,\kappa)=\mathrm{min}\{1,
w(e,u^{\mathrm{final}},\kappa)/w(e,u^{\mathrm{initial}},\kappa)\}\,$.
Eqs.(\ref{FD1}--\ref{FD3}) suggest that
$\kappa=\langle\hat\beta\rangle_e$ maximizes $p(e,\kappa)\,$ (a short
Metropolis run provides a first $\kappa$ estimate). We obtain $\langle
p(e,\kappa)\rangle_e>0.99$ for $e\leq e^\mathrm{d}$, and still
$\langle p(e,\kappa)\rangle_{e=e^\mathrm{o}}>0.78$.

We simulated the ($Q\!=\!10$,$D\!=\!2$) Potts model~\cite{BAXTER}, for
$L=32,64,128,256,512$ and $1024$, sampling $\langle\hat\beta\rangle_e$
at 30 points evenly distributed in $-1.41666\leq e\leq-0.45\,.$ For
$L=512$, we made 15 extra simulations to resolve the narrow spinodal
peaks (26 extra points for $L=1024$).  Our Elementary Monte Carlo Step
(EMCS) was: $\mathrm{max}\{10,N/(\langle {\cal N}\rangle_e \langle
p(e,\kappa) \rangle_e)\}$ cluster-flip attempts (${\cal N}$: number of
spins in the traced cluster; it is of order one at $e^\mathrm{d}$ and
of order $N$ at $e^\mathrm{o}$). So, every EMCS we flip at least $N$
spins.  For each $e$, we performed $2\times 10^6$ EMCS, dropping the
first $10\%$ for thermalization.  A similar computation was carried
out for the ($Q\!=\!4$,$D\!=\!3$) Potts model~\cite{JANKEQ4} (for
details see Table~\ref{MITABLA} and~\cite{INPREPARATION}).


Our $\langle\hat\beta\rangle_e$ in $D\!=\!2$ is shown in
Fig.~\ref{FIG1}. Data reweigthing (\ref{FD3}) was used only to
reconstruct the narrow spinodal peaks.  To find the roots of
$\langle\hat\beta\rangle_e=\beta$, or to calculate the integrals in
Eqs.~(\ref{MAXWELL},\ref{SIGMAEQ}), we interpolated
$\langle\hat\beta\rangle_e$ using a cubic spline \footnote{Not the so
called natural spline. We fixed the derivative at the first(last) $e$
value, from a 3 points parabolic fit.}.  For $L\geq 512$ the
strip-droplet transitions produce two ``jumps'' in
$\langle\hat\beta\rangle_e$, causing oscillations in the interpolation
(Gibbs phenomenon), cured by either of two interpolation schemes, see
Fig.~\ref{FIG1}.

We obtain $\beta_\mathrm{c}^L\,$, $\Sigma^L\,$,
$e^\mathrm{o}_L(\beta_\mathrm{c}^L)\,$,
$e^\mathrm{d}_L(\beta_\mathrm{c}^L)\,$,
$C_L(e^\mathrm{o}_L(\beta_\mathrm{c}^L))\,$ and
$C_L(e^\mathrm{d}_L(\beta_\mathrm{c}^L))\,$ from the interpolation of
$\langle\hat\beta\rangle_e$, and of
$\mathrm{d}\langle\hat\beta\rangle_e/\mathrm{d}e$, see (\ref{FD2}).
Statistical errors are Jack-Knife's ~\cite{VICTORAMIT} (the $i$-th
block is obtained interpolating the $i$-th Jack-Knife blocks for
$\langle\hat\beta\rangle_e$).  There are also interpolation and
integration errors.  Fortunately, errors of order $\epsilon$ in
$e^\mathrm{o}_L(\beta_\mathrm{c}^L)$ or
$e^\mathrm{d}_L(\beta_\mathrm{c}^L)$ yield errors of order
$\epsilon^2$ in $\beta_\mathrm{c}^L$: the main error in
$\beta_\mathrm{c}^L$ is the quadrature error for
$\langle\hat\beta\rangle_e$ divided by the latent heat.  On the other
hand, $e^*_L(\beta_\mathrm{c}^L)$ is near to the droplet-strip
transition, and an error on it does have an impact on $\Sigma_L$.


In Table~\ref{MITABLA} are our results for ($D\!=\!2,Q\!=\!10$) and the known
large $L$ limits.  A fit for $c$ in
$\beta_\mathrm{c}^\infty-\beta_\mathrm{c}^L=c/L^D$~\cite{BORGS-KOTECKY}
is unacceptable for $L\geq 32$ ($\chi^D/\mathrm{d.o.f.}=14.32/4$), but
good for $L\geq 64$ ($\chi^D/\mathrm{d.o.f.}=1.77/3$): our accuracy
allows to detect subleading corrections.  A fit
$e^\mathrm{o}_L(\beta_\mathrm{c}^L)-e^\mathrm{o}=b_1/L^D$ works only
for $L\geq 256$ ($\chi^2/\mathrm{d.o.f.}=1.90/2$; for
$e^\mathrm{d}_L(\beta_\mathrm{c}^L)$ we get
$\chi^2/\mathrm{d.o.f.}=1.41/2$).  However, $\beta^{\mathrm{strip},L}$
(see caption to Table~\ref{MITABLA}) is compatible with
$\beta_\mathrm{c}^\infty$ for $L\geq 256$. Then, the simplest strategy
to get $\beta_\mathrm{c}^\infty$ and the latent heat is: (1) for $L$
large enough to display a strip phase, locate it with short runs, (2)
get $\beta^{\mathrm{strip},L}$ accurately, and (3) find the
leftmost(rightmost) root for
$\langle\hat\beta\rangle_e=\beta^{\mathrm{strip},L}$.

As for $\Sigma^L$, the inequality $\Sigma^\infty\leq
0.0473505$~\cite{JANKESIGMA} (equality under the hypothesis of complete
wetting) was violated by $1/L$ extrapolations performed with $L\leq
100$~\cite{MULTICANONICAL}. The reader may check (Table~\ref{MITABLA}) that
our data for $L\leq 256$ extrapolate above 0.0473505, but drop below for
$L\geq 512$.  Indeed, the consistency of our results for $\beta_\mathrm{c}^L$
imply that the integration error for $\langle\hat\beta\rangle_e$ is (at most)
$2\times 10^{-6}$ for $L=1024$. Hence, the integration error for $\Sigma_L$ is
at most $10^{-3}$. Adding it to the difference between the linear and the
step-like interpolation, Fig.~\ref{FIG1}, we obtain
$\Sigma^{L=1024}=0.043(2)$, which is slightly below 0.0473505.

As for ($Q\!=\!4$, $D\!=\!3$), see Table~\ref{MITABLA},
$\beta^{\mathrm{strip},L}$ has converged (within accuracy) for $L\geq
64$. Hence, our preferred estimate is
$\beta_\mathrm{c}^\infty\!=\!0.6286206(10)$, that may be compared with
Janke and Kapler's
$\beta_\mathrm{c}^\infty\!=\!0.62863(2)$~\cite{JANKEQ4}. Accordingly, we
find $e^\mathrm{o}(\beta^{\mathrm{strip},L})\!=\!-1.10537(4)$,
$e^\mathrm{d}(\beta^{\mathrm{strip},L})\!=\!-0.52291(2)$,
$C_L(e^\mathrm{o}(\beta^{\mathrm{strip},L}))\!=\!35.4(9)$, and $
C_L(e^\mathrm{d}(\beta^{\mathrm{strip},L}))\!=\!4.24(18)$. The reader will
note that $\beta_\mathrm{c}^{L\!=\!128}$ is far too high (for instance,
from the $\chi^2/\mathrm{d.o.f.}$ of the extrapolation
$\beta_\mathrm{c}^L\!=\!\beta_\mathrm{c}^\infty+cL^{-D}$). Therefore, the
integration error is $\sim 4\times 10^{-6}$ (larger than the
statistical one), which provides a bound for the error in the surface
tension: $\Sigma^{L=128}\!=\!0.0118(4)$.  This is compatible with
$\Sigma^{L=64}$, and provides a reasonable $\Sigma^\infty$.

\begin{table*}
\begin{center}
\begin{tabular}{|c|l|l|l|l|l|l|l|}
  \hline 
  $L^D$ & $\beta_\mathrm{c}^L$& $\Sigma^L$& 
  $-e^\mathrm{o}_L(\beta_\mathrm{c}^L)$&
  $-e^\mathrm{d}_L(\beta_\mathrm{c}^L)$&
  $-C_L(e^\mathrm{o}_L(\beta_\mathrm{c}^L)$&
  $-C_L(e^\mathrm{d}_L(\beta_\mathrm{c}^L))$& $\beta^{\mathrm{strip},L}$\\
  \hline
  $32^2$&1.423082(17)& 0.05174(9)& 1.3318(2)&0.5736(3),&5.13(13)&3.99(7)&1.42028(7)\\
  $64^2$& 1.425287(9) & 0.05024(11) & 1.3220(2)&0.5999(2)&6.44(17)&5.78(19)&1.42479(4)\\
  $128^2$&1.425859(7) & 0.049225(14) & 1.31676(16)& 0.61164(16)&7.4(3)&7.8(3)&1.42592(2)\\
  $256^2$& 1.426021(5) & 0.0488(2) & 1.31478(8) &0.61578(8)&8.0(3)&8.7(4)&1.42606(2)\\
  $512^{2(A)}$&1.426051(4) &0.0473(3) &1.31392(6) &0.61710(4)&8.6(4)&9.1(4)&1.426048(12)\\
  $512^{2(B)}$&1.426048(4) &0.0467(4) &1.31390(6) &0.61708(5)&8.6(4)&9.1(4)&1.426048(12)\\
  $1024^{2(A)}$&1.4260599(19)& 0.0430(3) & 1.31375(3) & 0.61748(3) &9.7(5) &8.7(4)&1.426066(9)\\
  $1024^{2(B)}$&1.4260600(18)& 0.0424(2) & 1.31375(3) & 0.61748(3) &9.7(5) &8.7(4)&1.426066(9)\\
  \hline
  $\infty^2$&1.4260624389\ldots& $\Sigma^\infty\leq 0.0473505$&1.3136366978\ldots & 0.6175872662\ldots &--- &---&1.4260624389\ldots\\
  \hline
  \hline
  $8^3$&0.627394(7)& 0.005591(10)&1.1553(7)&0.51412(12)&23.0(5)&3.856(16)&
  0.62625(4)\\
  $16^3$& 0.628440(3)& 0.007596(6) & 1.1189(4)&0.51818(5)&30.1(8)&3.620(13)&
  0.626687(15)\\
  $32^3$& 0.6285957(10)& 0.009824(6) & 1.10751(15) &0.522066(16)&34.2(9)&4.019(17)&
  0.627889(6)\\
  $64^3$&0.6286133(7) & 0.011557(6)& 1.10542(8) &0.522831(8)&33.2(9)&4.11(2)&
  0.628621(3)\\
  $128^{3(A)}$&0.6286237(5)&0.011778(7)&1.10548(3)&0.52293(2)&35.4(9)&4.25(17)&
  0.6286206(10)\\
  $128^{3(B)}$&0.6286239(5)&0.011674(9)&1.10549(2)&0.52293(2)&35.4(9)&4.25(17)&
  0.6286206(10)\\
  \hline
\end{tabular}
\end{center}
\caption{System size dependent estimates of the quantities
  characterizing the first order transition, as obtained for the
  $Q\!=\!10,D\!=\!2$ Potts model ({\bf top}) and $Q\!=\!4,D\!=\!3$
  ({\bf bottom}).  Errors are Jack-Knife's. Also shown is
  $\beta^\mathrm{strip,L}\!=\!\langle\hat\beta\rangle_{e=-0.95}$ (for
  $D\!=\!2$) or
  $\beta^\mathrm{strip,L}\!=\!\langle\hat\beta\rangle_{e=-0.764443}$
  (for $D\!=\!3$), in the strip phase. The $\infty^2$ row contains
  exact results~{\protect\cite{BAXTER}} and an
  inequality~{\protect\cite{JANKESIGMA}}, for $D\!=\!2$,
  $Q\!=\!10\,$. The results with superscript $A$($B$) were obtained
  with the linear(step-like) interpolation scheme, see
  Fig.~\ref{FIG1}.  }\label{MITABLA}
\end{table*}


We propose a microcanonical strategy for the Monte Carlo simulation of
first-order phase transitions. The method is demonstrated in the
standard benchmarks: the $Q\!=\!10$, $D\!=\!2$ Potts model (where we
compare with exact results), and the $Q\!=\!4$, $D\!=\!3$ Potts model.
For both, we obtain accurate results in systems with more than $10^6$
spins (preexisting methods handle $10^4$ spins). Envisaged
applications include first-order transitions with quenched
disorder~\cite{CARDY,JANKEQ4}, colloid
crystallization~\cite{COLOIDES}, peptide
aggregation~\cite{JANKEPEPTIDOS} and the condensation
transition~\cite{CONDENSATION}.

We thank for discussions L.~A.  Fernandez (who also helped with
figures and C code), L.~G.  Macdowell, W. Janke, G. Parisi and P.
Verrocchio, as well as BIFI and the RTN3 collaboration for computer
time.  We were partly supported by BSCH---UCM and by MEC (Spain)
through contracts BFM2003-08532, FIS2004-05073.


\end{document}